\documentclass[twocolumn,prb,showpacs]{revtex4-1}
\usepackage{graphicx}

\newcommand{\crossprod}{\times}

\newcommand{\breakeq}{\nonumber \\ &&}
\newcommand{\cE}{\mathcal{E}}
\newcommand{\taubar}{\overline{\tau}}

\newcommand{\unit}[1]{\mathrm {\,#1}}

\begin{document}
%\draft

\title{
   Electromagnetic Green's function for layered systems:
   Applications to nanohole interactions in thin metal films
}

\author{Peter Johansson}
\email{petjo@chalmers.se}

\affiliation{
School of Science and Technology, University of \"Orebro, S--701 82
\"Orebro, Sweden 
\\
and
Department of Applied Physics, Chalmers University of Technology,
S--412\,96 G\"oteborg, Sweden
}

\date{\today}

\begin{abstract}
We derive expressions for the electromagnetic Green's function for a
layered system using a transfer matrix technique. The expressions we
arrive at makes it possible to study symmetry properties of the Green's
function, such as reciprocity symmetry, and the long-range properties of
the Green's function which involves plasmon waves as well as boundary
waves, also known as Norton waves.
We apply the method by calculating the light scattering cross section
off a chain of nanoholes in a thin Au film. The results highlight the
importance of nanohole interactions mediated by surface plasmon
propagating along the chain of holes.
\end{abstract}
%PACS
%68.37.Uv   Near-field scanning microscopy and spectroscopy
%78.68.+m   Optical properties of surfaces
%78.67.-n   Optical properties of nanoscale materials and structures
%73.20.Mf Collective excitations (including excitons, polarons,
%plasmons and other charge-density excitations)
%73.22.Lp    Collective excitations  (of nanoscale materials)
%\pacs{68.37.Uv,78.68.+m,78.67.-n,73.20.Mf,73.22.Lp}
\pacs{
42.25.Bs,  % Wave propagation
42.25.Fx,  % Diffraction and scattering
78.67.-n,  % Optical properties of nanoscale materials and structures
73.20.Mf   % Collective excitations (including excitons, polarons,
}

\maketitle
%%%%%%%%%%%%%%%%%%%%%%%%%%%%%%%%%%%%

% BEGIN_MUL_COMM (sed tag)
%\begin{multicols}{2}
% END_MUL_COMM (sed tag)

\section{Introduction}
\label{sec:introduction}

The research field of plasmonics\cite{Barnes:03} 
has seen an enormous development over
the last decades, both experimentally and theoretically. Examples of the
aspects studied include enhanced light emission under various
circumstances,\cite{Gimzewski:1989,*Berndt:1991,*Qiu:Ho:2003,
*Dong:2004,*Schneider:Berndt:2010} 
enhanced spectroscopies such as surface-enhanced Raman scattering
(SERS),\cite{Nie:Emory:1997,*Kneipp:1997,*Xu:Kall:1999,*Michaels:Brus:1999} 
extraordinary transmission of light\cite{Ebbesen:98}, and biosensing
applications.\cite{Dahlin:Hook:2005}

A corresponding development has also taken place on the theory side and
a
variety of methods are used to solve theoretical problems in
plasmonics. These include exact methods that apply for certain
geometries such as Mie
theory\cite{Waterman:71,Abajo:99,Xu:Mie:03,Johansson:05} 
for problems with spherical symmetry, but
in most situations methods that make more extensive use of numerical
calculations are needed.
The finite-difference in time-domain method 
(FDTD)\cite{Yee:FDTD:66,Chan:FDTD:95,Ward:98,Oubre:FDTD:04}
is one such method that has grown in popularity in
recent years, the discrete-dipole approximation (DDA) 
method\cite{Draine:94,Purcell:73}
and Green's function (GF)
method\cite{Martin:JOSA:94,Martin:GF:95,Paulus:00,Nikitin:2008,
Jung:2008,Simsek:2010} 
are two other
methods that are often used. Of these the DDA method has a somewhat
longer history and probably a bigger user base. The Green's function
method on the other hand can be more flexible in certain situations.

In this paper we will present a calculation of the Green's function for
a layered material. In particular this makes it possible to study
scattering off embedded inclusions such as nanoholes in metal films.
\cite{Tomas:hole:07,Alaverdyan:hole:natphys:07,Abajo:RMP07,Park:ACSNano:08,
Sepulveda:OptEx:2008, Alegret:NJP:2008, Lee:Nordlander:09,
Vesseur:Abajo:2009}
Paulus {\it et al.} presented a derivation of the Green's function
(Green's tensor) for a layered system
in Ref.\ \onlinecite{Paulus:00}. The present derivation follows the same basic
ideas, but we derive rather elegant, explicit expressions for the Green's 
function that
only involve a single transfer matrix recursion relation, and which
makes it possible to explicitly demonstrate various symmetry properties
of the Green's function such as reciprocity symmetry. We also study the
analytic properties of the Green's function in Fourier space and show
how this effects the long-range properties of the Green's function which
for metallic films are dominated by plasmon polaritons for distances
typically in the range of 100 nm to 10 $\unit{\mu m}$ and for even
larger distance the dominating contribution comes from a boundary wave.

We will illustrate the Green's function method by calculating 
scattering cross sections for light off nanohole systems in thin metal films.
The nanoholes of these systems typically have 
diameters that range from 50 to 100 nm in a thin Au film of thickness 20 nm.
The optical properties of nanohole system have attracted intense
interest
for over a decade now in the context of extraordinary transmission
through an array of nanoholes discovered by Ebbesen and 
coworkers,\cite{Ebbesen:98} and studied by various theoretical methods;
\cite{Porto:Pendry:PRL1999,MartinMoreno:2001}
for a couple of recent reviews on this subject see Refs.\
\onlinecite{GarciaVidal:10, Gordon:Brolo:2010}.
But nanohole systems are also studied in connection with
biosensing applications, since they at the same time can act 
as capturing centers for biomolecules and light
scatterers whose properties are modulated by the presence of these
molecules.
The basic optical scattering properties of individual nanoholes and
chains of nanoholes in thin metal (Au films)
have been studied by K\"all and co-workers and the results
show signs of strong hole-hole
interactions.\cite{Alaverdyan:hole:natphys:07} 
Here we present theoretical results for the scattering cross section off
multi-hole systems
that are in good agreement with the experimental ones. 
The theoretical results combined with an analysis of the
behavior of the Green's function shows that the hole-hole interaction
affecting the light scattering is
to a large extent mediated by surface (interface) plasmons. The special nature of the
plasmons means that there is a strong relation between polarization and
propagation direction; hole-hole interactions are much
stronger in the case when the electric field is polarized along
the axis through the hole centers than when the polarization is
perpendicular to the chain axis. 

The rest of the paper is organized in the following way. In section
\ref{sec:general:green} we give a brief overview of the Green's function
method.
Section \ref{sec:derivation} details our calculation of the
Green's function for a layered background through a transfer-matrix
method and we also show how a number of physical quantities can be derived
from the Green's function.
Section \ref{sec:analytic} focuses
on the analytic properties of the Green's function in wave vector space
and its consequences for the long range behavior in real space.
Section \ref{sec:solution} gives a brief description of the numerical
solution of the integral equation determining the electric field in the
scatterers.
In Sec.\ \ref{sec:holes} we apply the method to a
study of the optical properties of nanoholes in a thin metal film, and
the paper is summarized in Sec.\ \ref{sec:summary}.

\section{Basic treatment of the scattering problem}
\label{sec:general:green}

We consider first a situation where all of space is filled with a
material with dielectric function $\varepsilon_B$, corresponding to a
wave number
\begin{equation}
   k_B = \sqrt{\varepsilon_B} k_0 = \sqrt{\varepsilon_B} (\omega/c),
\end{equation}
where $k_0$ and $c$ are the wave number and speed of light in vacuum,
respectively, and $\omega$ the angular frequency of electric and
magnetic fields.
The task at hand is to solve Maxwell's equations which assuming all
fields have a $e^{-i\omega t}$ time dependence, read
\begin{equation}
   \nabla \crossprod \vec{E} 
   =
   i \omega \vec{B},
\label{eq:rotE}
\end{equation}
\begin{equation}
   \nabla \crossprod B 
   =
   \mu_0 \vec{j}
   -
   i \omega
   \mu_0 \varepsilon_0 \varepsilon_B 
   \vec{E},
\end{equation}
along with
\begin{equation}
   \nabla \cdot \vec{B} = 0, 
   \ \ \mathrm{and}\  \ 
   \nabla \cdot \vec{D} = \rho.
\end{equation}

The solution for the electric field can in this case be
written as the sum of a source term that depends on the current at the
field point $\vec{r}$ and another term that through a Green's function
takes into account the effects of the currents everywhere
else,\cite{Yaghjian:80}
\begin{equation}
   \vec{E} (\vec{r})
   =
   \frac{\tensor{L}\cdot\vec{j}(\vec{r})}{i\omega\varepsilon_0\varepsilon_B}
   +
   i \omega \mu_0
   \int_{V_j-V_{\delta}}
   \tensor{G}_{h}(\vec{r}, \vec{r}')
   \cdot
   \vec{j}(\vec{r}')
   d^3r'.
\end{equation}
Here $\tensor{L}$ is a tensor which depends on the shape of the excluded
volume $V_{\delta}$. In the most common cases where the excluded volume
is cubic or spherical in shape $\tensor{L}$ is diagonal and each of
its three elements have the value 1/3. The Green's function is given by
\begin{equation}
   \tensor{G}_h(\vec{r},\vec{r'})
   =
   \left[\tensor{1} + \frac{\nabla\nabla}{k_B^2} \right]
   G^{\mathrm{s}} (\vec{r}, \vec{r'})
   =
   \left[\tensor{1} + \frac{\nabla\nabla}{k_B^2} \right]
   \frac{e^{ik_B|\vec{r}-\vec{r'}|}}{4\pi |\vec{r}-\vec{r'}|},
\label{eq:Ghom}
\end{equation}
where $G^{\mathrm{s}}$ is the Green's function to the scalar Helmholtz
equation, and thus satisfies
\begin{equation}
   (\nabla^2 + k_B^2) G^{\mathrm{s}}(\vec{r},\vec{r'})
   =
   - \delta^{(3)}(\vec{r}-\vec{r}').
\end{equation}
More explicitly we have 
\begin{eqnarray}
   \tensor{G}_h(\vec{R})
   &&
   =
   \left(
      \tensor{1} 
      +
      \frac{ik_BR-1}{k_B^2R^2}
      \tensor{1}
   \right.
   \breakeq
   \left.
      +\frac{3-3ik_BR-k_B^2R^2}{k_B^2R^4}
      \vec{R}\otimes\vec{R}
   \right)
   \frac{e^{ik_BR}}{4\pi R},
\label{eq:Ghexplicit}
\end{eqnarray}
where $\vec{R} = \vec{r}-\vec{r'}$, and $\otimes$ denotes a dyadic
product.

In case the dielectric function is not
constant in space, there is a modification of the second of the
Maxwell's equations 
\begin{equation}
   \nabla \crossprod \vec{B}
   =
   \mu_0 \vec{j}
   -
   \frac{i \omega \varepsilon_B}{c^2} \vec{E}
   - 
   \frac{i \omega (\varepsilon_{\mathrm{rel}}-\varepsilon_B)}{c^2} \vec{E}, 
\end{equation}
where the last term is new compared with the case of a homogeneous
medium, and $\varepsilon_{\mathrm{rel}}$ can vary in space in an
essentially arbitrary way. We rewrite this as 
\begin{equation}
   \nabla \crossprod \vec{B}
   =
   \mu_0 \vec{j}_{\mathrm{tot}}
   -
   \frac{i \omega \varepsilon_{B}}{c^2} \vec{E}
\end{equation}
where the total current due to both external sources and scattering  is
\begin{equation}
   \vec{j}_{\mathrm{tot}}
   =
   \vec{j} + \vec{j}_{\mathrm{scatt}}
   =
   \vec{j}
   -
   \frac{1}{\mu_0}
   \frac{i\omega(\varepsilon_{\mathrm{rel}}-\varepsilon_B)}{c^2}
   \vec{E}.
\end{equation}

By letting $\vec{j}_{\mathrm{tot}}$ take the place of $\vec{j}$ in the
solution we get
\begin{eqnarray}
   \vec{E} (\vec{r})
   &&
   =
   \frac{\tensor{L}\cdot\vec{j}(\vec{r})}{i\omega\varepsilon_0\varepsilon_B}
   -
   \frac{\Delta \varepsilon(\vec{r)}}{\varepsilon_B} 
   \tensor{L} \cdot \vec{E}(\vec{r})
   +
   \int_{V_j-V_{\delta}}
   \tensor{G}_{h}(\vec{r}, \vec{r}')
   \breakeq
   \times
   \left(
      i \omega \mu_0
      \vec{j}(\vec{r}')
      +
      k_0^2 \Delta \varepsilon(\vec{r'}) \vec{E}(\vec{r'})
   \right)
   d^3r'.
\label{eq:lipp}
\end{eqnarray}
We will next show how this expression, in particular the Green's
function,
can be generalized to deal with a layered system where the
background dielectric function $\varepsilon_B(\vec{r})$
varies stepwise along one direction ($z$) in
space, and $\Delta \varepsilon=\varepsilon_{\mathrm{rel}}-\varepsilon_B$ 
then describes further variations of
the dielectric function due to scatterers.

\section{Green's function for a layered structure}
\label{sec:derivation}

\subsection{Formulation in terms of 2D Fourier transform}

The Green's function for a homogeneous background given above in Eq.\
(\ref{eq:Ghom}) can be written in terms of a Fourier integral as
\begin{equation}
   \tensor{G}_h(\vec{R})
   =
   \int
   \frac{d^3q}{(2\pi)^3k_B^2} 
   \frac{k_B^2\tensor{1}  - \vec{q}\otimes \vec{q}}{q^2 - k_B^2}
   e^{i\vec{q}\cdot\vec{R}}.
\end{equation}
This expression becomes more useful in handling layered structures if we
integrate out the $q_z$ variable, which yields\cite{Paulus:00}
\begin{equation}
   \tensor{G}_h(\vec{R})
   =
   - \frac{\hat{z}\otimes\hat{z}}{k_B^2}
   \delta^{(3)}(\vec{R})
   +
   \int
   \frac{d^2q}{(2\pi)^2} 
   \tensor{G}_h(\vec{q}_{\|}, z)
   e^{i\vec{q}_{\|}\cdot\vec{r}_{\|}}.
\label{eq:Fourier}
\end{equation}
The notation $\tensor{G}_h$ indicates that we are still dealing with the
Green's function for a homogeneous background, but the formal
generalization of Eq.\ (\ref{eq:Fourier}) to $\tensor{G}$ valid for a
layered background is straightforward.
The $\delta$-function term is the result of a subtract-add operation
necessary to render the contour integral convergent. However, in the
following we will not explicitly deal with the singular behavior of
$\tensor{G}$ when $\vec{r}\to\vec{r}'$, so we will therefore leave out
this term in the rest of the calculation.
The 2D Fourier transform (FT) of the Green's
function in Eq.\ (\ref{eq:Fourier}) is 
\begin{equation}
   \tensor{G}_h(\vec{q}_{\|}, z, z')
   =
   \frac{i}{2p} 
   \left(
      \tensor{1}
      -
      \frac{\vec{q}^{\,\tau} \otimes \vec{q}^{\,\tau}}{k_B^2}
   \right)
   e^{ip|z-z'|},
\label{eq:Gq2d}
\end{equation}
where $p$ stands for the absolute value of the $z$ component of the wave
vector $\vec{q}$ (originating from the residue at the pole in the
contour integration) 
\begin{equation}
   p 
   =
   \sqrt{k_B^2 - |\vec{q}_{\|}|^2}.
\label{eq:pdef}
\end{equation}
The square root function in Eq.\ (\ref{eq:pdef}) should, to give
physical results in the form of outgoing, damped waves, be evaluated
with the branch cut along the positive real axis of the argument.
The superscript $\tau$ on the wave vector
$\vec{q}^{\,\tau}$ indicates the direction of propagation of the waves
(called {\em primary propagation direction} in the following) $\tau=+1$,
or just $\tau=+$, when
$z>z'$ and $\tau=-1$ when $z<z'$. For the wave vectors we have 
\begin{equation}
   \vec{q}^{\pm}
   =
   \vec{q}_{\|} \pm p \hat{z}.
   =
   (
      (q_{\|}/k_B) \cos{\phi_q},
      (q_{\|}/k_B) \sin{\phi_q},
      \pm p/k_B
   ),
\end{equation}
where $q_{\|}= |\vec{q}_{\|}|$.
The corresponding unit vector, 
\begin{equation}
   \hat{q}^{\,\tau} = \vec{q}^{\,\tau}/ k_B,
\label{eq:qhat}
\end{equation}
together with the unit polarization vectors for s polarization
\begin{equation}
   \hat{s}^{\pm} 
   =
   \frac{\hat{z}\crossprod\hat{q}^{\pm}}{|\hat{z}\crossprod\hat{q}^{\pm}|}
   =
   (-\sin{\phi_q}, \cos{\phi_q}, 0),
\end{equation}
and p polarization
\begin{equation}
   \hat{p}^{\pm}
   =
   \hat{s}^{\pm}
   \crossprod
   \hat{q}^{\pm}
   =
   \left(
      \pm\frac{p}{k_B} \cos{\phi_q}, 
      \pm\frac{p}{k_B} \sin{\phi_q}, 
      - \frac{q_{\|}}{k_B}
   \right),
\end{equation}
form an orthonormal basis. The unit tensor therefore can be written
\begin{equation}
   \tensor{1}
   =
   \hat{q}^{\tau} \otimes \hat{q}^{\tau}
   +
   \hat{p}^{\tau}\otimes\hat{p}^{\tau}
   +
   \hat{s}^{\tau}\otimes\hat{s}^{\tau}.
\label{eq:unit_tensor}
\end{equation}
Using Eqs.\ (\ref{eq:qhat}) and (\ref{eq:unit_tensor}) we can rewrite
the Green's function Fourier transform in Eq.\ (\ref{eq:Gq2d}) as
\begin{equation}
   \tensor{G}_h(\vec{q}_{\|}, z, z')
   =
   \frac{i}{2p} 
   \left(
      \hat{p}^{\tau}\otimes\hat{p}^{\tau} 
      +
      \hat{s}^{\tau}\otimes\hat{s}^{\tau}
   \right)
   e^{i\tau p(z-z')}.
\label{eq:Gqhom}
\end{equation}

% It should be noted that, with the exception of $\hat{s}$ which only
% depends on the in-plane direction of $\vec{q}$, the unit vectors as well
% as $k_B$ and $p$ are layer-dependent. 

As a first step towards generalizing Eq.\ (\ref{eq:Gqhom}) to a
situation with a layered background, we conclude that a particular element
of the tensor can be written
\begin{eqnarray}
   &&
   G_{h,\alpha\beta}(\vec{q}_{\|}, z, z')
   = 
   \hat{\alpha} 
   \cdot
   \vec{\cE}
   =
   \breakeq
   =
   \hat{\alpha}
   \cdot
   \left[
   (
      \hat{p}^{\tau} A^{\tau,p} 
      +
      \hat{s}^{\tau} A^{\tau,s} 
   )
   e^{i\tau p(z-z')}
   \right], 
\label{eq:gkhom}
\end{eqnarray}
where $\vec{\cE}$ is a vector field proportional to an electric field
generated by the source.
The wave amplitudes are found by projecting the source unit vector
$\hat{\beta}$ onto the p and s unit vectors, thus
\begin{equation}
   A^{\tau,p} 
   =
   \frac{i}{2p}
   \,
   \hat{p}^{\tau} 
   \cdot
   \hat{\beta}
   \ \ \ \ \mathrm{and} \ \ \ \ 
   A^{\tau,s} 
   =
   \frac{i}{2p}
   \,
   \hat{s}^{\tau} 
   \cdot
   \hat{\beta}.
\end{equation}

\subsection{Generalization to a layered material}

When we turn to a layered material the source will still generate
outgoing plane waves just like the expression in Eq.\ (\ref{eq:gkhom})
indicates, however, now there will also be other waves reflected and
transmitted at the different interfaces. 

\begin{figure}[tb]
   \includegraphics[angle=0,width=8.0 cm]{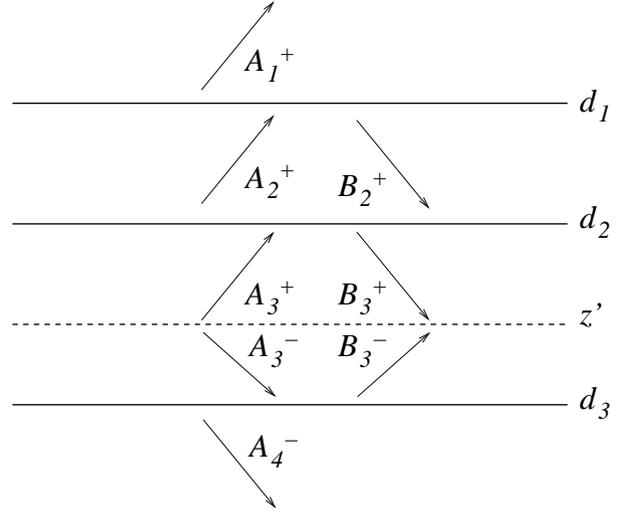}
   \caption{
      Illustration of the plane waves generated in the different layers
      of a four-layer system when the source is placed at $z=z'$.
   }
\label{fig:waves}
\end{figure}

% Introduces A and B
This is illustrated in 
Fig.\ \ref{fig:waves}.
for a system with four layers and the
source placed in layer 3. In layer 3, there are waves going upwards and
downwards on both sides of the source, in layer 2 there are also waves
propagating in both directions, but in the two outermost layers there
are only outgoing waves, propagating upwards in layer 1 and downwards in
layer 4.  The vector field $\vec{\cE}$ we introduced in Eq.\
(\ref{eq:gkhom})
takes the
generalized form
\begin{eqnarray}
   &&
   \vec{\cE} =
   \left[
   \left(
      \hat{p}_l^{\tau} 
      A_l^{\tau,p}(z_{0,l}) 
      + 
      \hat{s}
      A_l^{\tau,s}(z_{0,l}) 
   \right)
   e^{i\tau p_l (z-z_{0,l})}
   \right.
   \breakeq
   \left.
      +
   \left(
      \hat{p}_l^{\taubar} 
      B_l^{\tau,p}(z_{0,l}) 
      + 
      \hat{s}
      B_l^{\tau,s}(z_{0,l}) 
   \right)
   e^{-i\tau p_l (z-z_{0,l})}
   \right],
\label{eq:vectorfield}
\end{eqnarray}
in layer $l$.
Here $\taubar$ is the opposite direction of $\tau$, $-\tau$, and the
unit vectors for p
polarization as well as $p$ depends on the layer number $l$ since the
background wave vector magnitude $k_B$ varies from layer to layer.
The wave amplitudes 
$A_l^{\tau,\sigma}(z_{0,l})$ 
for outgoing waves propagating away from $z=z'$
and
$B_l^{\tau,\sigma}(z_{0,l})$ 
for returning waves propagating towards $z=z'$
in Eq.\ (\ref{eq:vectorfield}) depend on the layer number $l$, primary
propagation direction $\tau$, and polarization $\sigma$ (p or s). The offset
points $z_{0,l}$ are local origins for the plane wave
exponentials, which can be moved around provided, of course, that the
wave amplitudes are adjusted accordingly.
The standard choice for 
$z_{0,l}$, in particular in a numerical implementation, 
is to use 
the bottom of all the layers above the source, and the top of all the
layers below the source. In Fig.\ \ref{fig:waves} this means that
$z_{0,1}=d_1$,
$z_{0,2}=d_2$,
$z_{0,3}=z'$,
and 
$z_{0,4}=d_3$.

% New stuff
We need to determine the wave amplitudes
$A_l^{\tau,\sigma}(z_{0,l})$ 
and
$B_l^{\tau,\sigma}(z_{0,l})$, 
something we will do in two steps. First we view the stack of layers
as made up of two independent parts, one above the source plane $z=z'$,
and one below. We introduce relative wave amplitudes 
$a_l^{\tau,\sigma}(z_{0,l})$ 
and
$b_l^{\tau,\sigma}(z_{0,l})$, corresponding to the actual
amplitudes
$A_l^{\tau,\sigma}(z_{0,l})$ 
and 
$B_l^{\tau,\sigma}(z_{0,l})$. 
The actual amplitudes above the source are found from the relative ones as 
\begin{equation}
   A_l^{+}(z'') = \frac{a_l^+(z'')}{a_{l'}^+(z')} A_{l'}^+(z'),
   \ \ \ \
   B_l^{+}(z'') = \frac{b_l^+(z'')}{a_{l'}^+(z')} A_{l'}^+(z'),
\label{eq:rel2actplus}
\end{equation}
where $l'$ denotes the source layer and $z''$ is any $z$ coordinate.
In the same way, below the source
\begin{equation}
   A_l^{-}(z'') = \frac{a_l^-(z'')}{a_{l'}^-(z')} A_{l'}^-(z'),
   \ \ \ \
   B_l^{-}(z'') = \frac{b_l^-(z'')}{a_{l'}^-(z')} A_{l'}^-(z').
\label{eq:rel2actminus}
\end{equation}
The relative amplitudes can be determined by a transfer-matrix
calculation using the fact that there are only outgoing waves in the
outermost layers. 
But Eqs.\ (\ref{eq:rel2actplus}) and (\ref{eq:rel2actminus}) show that
the actual wave amplitudes $A_{l'}^+$ and $A_{l'}^-$ in the source layer
play the role of driving forces for all the waves above and below the
source, respectively, and still have to be calculated independently.
This is done in the second step of our calculation, by a detailed
investigation of the situation in the source layer.

\subsection{Transfer matrix calculation}
\label{subsec:transfer} 

The calculation of the relative amplitudes uses the fact that
there are no returning waves in the outermost layers, thus
\begin{equation}
   B_1^+ = b_1^+ = 0, 
   \ \ \ \mathrm{and} \ \ \ 
   B_N^- = b_N^- = 0. 
\end{equation}
We set the relative
amplitudes of the outgoing waves in the outermost layers, i.e. $a_1^+$
and $a_N^-$, to 1, 
\begin{equation}
   a_1^+(z_+) = 1, 
   \ \ \ \mathrm{and} \ \ \ 
   a_N^-(z_-) = 1.
\end{equation}
The coordinates $z_+$ and $z_-$ lie above ($z_+$) and below ($z_-$) all
interfaces as well as the source and field points, respectively.

The
remaining relative amplitudes can then be determined by recursion, applying the
Fresnel formula at each interface and adjusting the amplitudes by
exponentials to account for the propagation through intermediate layers.
For a general $z$ coordinate $z''$ we get
\begin{eqnarray}
   &&
   \left[
      \begin{array}{c}
      a_{l}^{+,\sigma}(z'') \\
      b_{l}^{+,\sigma}(z'') \\
      \end{array}
   \right]
   =
   W^{+}(z'', z_+) 
   \left[
      \begin{array}{c}
      1 \\
      0 \\
      \end{array}
   \right]
   \breakeq
   =
   S_{l}^{+}(z'', d_{l-1}) 
   T_{l,l-1}^{+,\sigma} 
   \cdots
   T_{2,1}^{+,\sigma} 
   S_{1}^{+}(d_1, z_+)
   \left[
      \begin{array}{c}
      1 \\
      0 \\
      \end{array}
   \right]
\label{eq:Wplusdef}
\end{eqnarray}
above the source, and
\begin{eqnarray}
   &&
   \left[
      \begin{array}{c}
      a_{l}^{-,\sigma}(z'') \\
      b_{l}^{-,\sigma}(z'') \\
      \end{array}
   \right]
   =
   W^{-}(z'', z_-) 
   \left[
      \begin{array}{c}
      1 \\
      0 \\
      \end{array}
   \right]
   =
   \breakeq
   =
   S_{l}^{-}(z'', d_l)
   T_{l,l+1}^{-,\sigma} 
   \cdots
   T_{N-1,N}^{-,\sigma} 
   S_{N}^{-}(d_{N-1}, z_-)
   \left[
      \begin{array}{c}
      1 \\
      0 \\
      \end{array}
   \right]
\label{eq:Wminusdef}
\end{eqnarray}
below the source. Here $W$ is a ``total'' transfer matrix built up by
factors $S$, related to the wave
propagation in the different layers, and $T$, describing
reflection and transmission at a particular interface.

%%%%%%%%%%%
The propagation in one single layer just yields exponential factors
multiplying the wave amplitudes. We have
\begin{equation}
   \left[
      \begin{array}{c}
      a_{l}^{\tau,\sigma}(z) \\
      b_{l}^{\tau,\sigma}(z) \\
      \end{array}
   \right]
   =
   S_{l}^{\tau}(z,z'')
   \left[
      \begin{array}{c}
	 a_{l}^{\tau,\sigma}(z'') \\
	 b_{l}^{\tau,\sigma}(z'') \\
      \end{array}
   \right],
\end{equation}
where $\sigma$ denotes a polarization (s or p) and where
\begin{equation}
   S_{l}^{\pm}(z,z'')
   =
   \left[
      \begin{array}{cc}
      e^{\pm i p_l(z-z'')} & 0 \\
      0 & e^{\mp i p_l(z-z'')} \\
      \end{array}
   \right].
\end{equation}

The cross-interface transfer matrices $T$ relate the wave coefficients
on opposite sides of an interface $z=d$, separating layers $l$ and $n$
(where $l=n\pm 1$),
to each other
\begin{equation}
   \left[
      \begin{array}{c}
      a_{l}^{\tau,\sigma}(d) \\
      b_{l}^{\tau,\sigma}(d) \\
      \end{array}
   \right]
   =
   T_{ln}^{\tau,\sigma}
   \left[
      \begin{array}{c}
	 a_{n}^{\tau,\sigma}(d) \\
	 b_{n}^{\tau,\sigma}(d) \\
      \end{array}
   \right].
\end{equation}
They can be evaluated by using the
Fresnel formulae for s and p polarized waves.
We express the result in terms of 
the reflection amplitudes for $s$ and $p$ polarized waves, respectively, 
incident from the material in layer $l$ onto material $n$ in case
this is the only interface,
\begin{equation}
   f_{ln}^s = \frac{p_{l} - p_{n}} {p_{l} + p_{n}}
\ \ \ \mathrm{and} \ \ \ 
   f_{ln}^p = \frac{\varepsilon_{n}p_{l} - \varepsilon_{l}p_{n}} 
   {\varepsilon_{n}p_{l} + \varepsilon_{l}p_{n}}.
\label{eq:Fresnelf}
\end{equation}
These quantities 
depend on the dielectric functions $\varepsilon_l$ and
$\varepsilon_n$ and wave vector $z$ components $p_l$ and $p_n$ of the
two layers.
The $T$ matrices 
are found after some algebra which also involves the amplitude of the
transmitted wave. We get
\begin{equation}
   T_{ln}^{\tau s} 
   =
   \frac{1}{1+f_{ln}^s}
   \left[
      \begin{array}{cc}
      1 & f_{ln}^s \\
      f_{ln}^s &  1
      \end{array}
   \right]
\end{equation}
for s polarization,
and 
\begin{equation}
   T_{ln}^{\tau p}
   =
   \frac{k_lp_n}{k_np_l}
   \frac{1}{1-f_{ln}^p}
   \left[
      \begin{array}{cc}
      1 & f_{ln}^p \\
      f_{ln}^p &  1
      \end{array}
   \right]
\end{equation}
for p polarization.

\subsection{Calculation of the primary wave}

Using the scheme outlined in Sec.\ \ref{subsec:transfer} we can calculate all
the relative wave amplitudes we need. Now it remains to find the primary
wave amplitudes $A_{l'}^{\pm,\sigma}$ in the source layer. 
We have already seen that in the case of a homogeneous material we have 
\begin{equation}
   A_{l'}^{\tau,p} (z')
   =
   \frac{i}{2p_{l'}}
   \,
   \hat{p}^{\tau} 
   \cdot
   \hat{\beta}
   \ \ \ \ \mathrm{and} \ \ \ \ 
   A_{l'}^{\tau,s} (z')
   =
   \frac{i}{2p_{l'}}
   \,
   \hat{s}^{\tau} 
   \cdot
   \hat{\beta}.
\end{equation}
In the present case we have to add the wave reflected off the
``opposite'' interface of the source layer to each of these expressions. 
Thus, the amplitude
of the wave propagating upwards has a direct contribution from the
source, and one contribution from the interface below the source, and
vice versa for the wave propagating downwards. By introducing the
response functions, i.e. the ratios between reflected and incident wave
amplitudes, 
\begin{equation}
   \chi^{\tau,\sigma}(z'')
   =
   \frac{b_{l}^{\tau,\sigma}(z'')}{a_{l}^{\tau,\sigma}(z'')}
   =
   \frac{B_{l}^{\tau,\sigma}(z'')}{A_{l}^{\tau,\sigma}(z'')}
\end{equation}
we can write (using $\hat{\sigma}$ as a general polarization vector,
$\hat{s}$ or $\hat{p}$)
each of these amplitudes as
\begin{equation}
   A_{l'}^{\tau,\sigma}(z')
   =
   \frac{i}{2p_{l'}}
   \,
   \hat{\sigma}_{l'}^{\tau} 
   \cdot
   \hat{\beta}
   + 
   \chi^{\taubar,\sigma}(z')
   A_{l'}^{\taubar,\sigma}(z'),
   \ \ \tau=\pm 1.
\end{equation}
This system of two equations 
has the solution
\begin{equation}
   A_{l'}^{\tau,\sigma}(z')
   =
   \frac
   {
      (
      \hat{\sigma}_{l'}^{\tau}
      +
      \hat{\sigma}_{l'}^{\taubar} \,
      \chi^{\taubar,\sigma}(z')
      )
      \cdot
      \hat{\beta}
   }
   {1 - \chi^{+,\sigma}(z') \chi^{-,\sigma}(z')}
   \,
   \frac{i}{2p_{l'}}
\end{equation}
in which the first term in the numerator is the direct wave from the
source, the second term is the wave reflected once off the opposite
interface, and the denominator accounts for repeated reflections off the
surrounding interfaces.
We can now calculate the Green's function by using this solution in
Eqs.\ (\ref{eq:vectorfield}), (\ref{eq:rel2actplus}), and
(\ref{eq:rel2actminus}).

\subsection{Results for the Fourier-space GF}

With a source pointing in the $\beta$ direction we can now write down
the result for the matrix element 
$G_{\alpha\beta}(\vec{q}_{\|},z, z') = \hat{\alpha}\cdot\cE$ in Fourier
space.
To keep the whole thing manageable we divide the Green's function
into one p part and one s part
\begin{equation}
   \tensor{G}(\vec{q}_{\|},z,z')
   =
   \tensor{G}^p(\vec{q}_{\|},z,z')
   +
   \tensor{G}^s(\vec{q}_{\|},z,z'),
\label{eq:Gqsum}
\end{equation}
with
\begin{widetext}
\begin{equation}
   \tensor{G}^p(\vec{q}_{\|},z,z')
   =
   \left[
      \hat{p}_l^{\tau} 
      \,
      \frac{
	 a_l^{\tau,p}(z_{0,l}) 
      }
      {
	 a_{l'}^{\tau,p}(z') 
      }
   e^{i\tau p_l (z-z_{0,l})}
      +
      \hat{p}_l^{\taubar} 
      \,
      \frac{
	 b_l^{\tau,p}(z_{0,l}) 
      }
      {
	 a_{l'}^{\tau,p}(z') 
      }
   e^{-i\tau p_l (z-z_{0,l})}
   \right]
   \otimes
   \frac
   {
      (\hat{p}_{l'}^{\tau}+\chi^{\taubar,p}(z')\hat{p}_{l'}^{\taubar})
   }
   {
      1 - \chi^{+,p}(z') \chi^{-,p}(z')
   }
   \,
   \frac{i}{2p_{l'}}
\label{eq:Gp}
\end{equation}
and 
\begin{equation}
   \tensor{G}^s(\vec{q}_{\|},z,z')
   =
   \left[
      \frac{
	 a_l^{\tau,s}(z_{0,l}) 
      }
      {
	 a_{l'}^{\tau,s}(z') 
      }
   e^{i\tau p_l (z-z_{0,l})}
      +
      \frac{
	 b_l^{\tau,s}(z_{0,l}) 
      }
      {
	 a_{l'}^{\tau,s}(z') 
      }
   e^{-i\tau p_l (z-z_{0,l})}
   \right]
   \hat{s}
   \otimes
   \hat{s}
   \,
   \frac
   {
      [1+\chi^{\taubar,s}(z')]
   }
   {
      1 - \chi^{+,s}(z') \chi^{-,s}(z')
   }
   \,
   \frac{i}{2p_{l'}}.
\label{eq:Gs}
\end{equation}
These expressions are a good starting point for a numerical implementation. 

For theoretical purposes it is, however, quite useful to express the
Green's functions in terms of the $W$ transfer matrices. By multiplying
the numerators and denominators in Eqs.\ (\ref{eq:Gp}) and (\ref{eq:Gs})
by 
$a_{l'}^{\tau,\sigma}(z') a_{l'}^{\taubar,\sigma}(z')$, and using that,
in view of Eqs.\ (\ref{eq:Wplusdef})  and (\ref{eq:Wminusdef}),
$a_{l''}^{\tau,\sigma}(z'')$ 
and 
$b_{l''}^{\tau,\sigma}(z'')$ 
are the
first column elements of $W^{\tau,\sigma}(z'',z_{\tau})$ we arrive at
\begin{equation}
   \tensor{G}^p(\vec{q}_{\|},z,z')
   =
   \frac{
      W_{11}^{\tau,p}(z,z_{\tau})
      \,
	 \left[
	    \hat{p}_{l}^{\tau}
	    +
	    \hat{p}_{l}^{\taubar}
	    \chi^{\tau,p}(z)
	 \right]
      \otimes
	 \left[
	    \hat{p}_{l'}^{\tau}
	    +
	    \hat{p}_{l'}^{\taubar}
	    \chi^{\taubar,p}(z')
	 \right]
      \,
      W_{11}^{\taubar,p}(z',z_{\taubar})
   }
   {
      D_{11}^p(z')
   }
\label{eq:Gpqformal}
\end{equation}
and 
\begin{equation}
   \tensor{G}^s(\vec{q}_{\|},z,z')
   =
   \frac{
      W_{11}^{\tau,s}(z,z_{\tau})
      \,
	 \left[
	    1
	    +
	    \chi^{\tau,s}(z)
	 \right]
	    \hat{s}
      \otimes
	    \hat{s}
	 \left[
	    1
	    +
	    \chi^{\taubar,s}(z')
	 \right]
      \,
      W_{11}^{\taubar,s}(z',z_{\taubar})
   }
   {
      D_{11}^s(z')
   },
\label{eq:Gsqformal}
\end{equation}
where 
\begin{equation}
   D_{11}^{\sigma}(z') 
   = 
   -2 i p_{l'} 
   \,
   \left[
      W_{11}^{+,\sigma}(z', z_+)
      W_{11}^{-,\sigma}(z', z_-)
      -
      W_{21}^{+,\sigma}(z', z_+)
      W_{21}^{-,\sigma}(z', z_-)
   \right]
\label{eq:D11long}
\end{equation}
\end{widetext}
is the $1,1$ element of the matrix
\begin{equation}
   D^{\sigma}(z') = 
   \left[
      W^{+,\sigma}(z',z_+)
   \right]^t
   (-2ip_{l'} \sigma_z)
   W^{-,\sigma}(z',z_-).
\label{eq:D11short}
\end{equation}
Here the superscript $t$ denotes matrix transposition, and $\sigma_z$ is
the Pauli matrix
\begin{equation}
   \sigma_z
   =
   \left[
      \begin{array}{cr}
      1 & 0 \\
      0 & -1 \\
      \end{array}
   \right].
\end{equation}
To summarize, Eqs.\ (\ref{eq:Gpqformal})--(\ref{eq:D11long}) show how the
Green's function can be uniquely expressed in terms of the transfer
matrices describing propagation from the outer layers of the system to
the source and field points $z'$ and $z$, respectively.

\subsection{Calculation of the real-space GF}

The layered system we are considering has cylindrical symmetry and it is
therefore quite natural to view both the Green's function in real space,
as well as its Fourier transform which has been at the center of our
attention so far, as functions of cylindrical coordinates
\begin{equation}
   \tensor{G}(\vec{r}, \vec{r}')
   \equiv
   \tensor{G}(\rho, \phi, z, z')
   \ \ \mathrm{and} \ \ 
   \tensor{G}(\vec{q}_{\|}, z, z')
   \equiv
   \tensor{G}(q_{\|}, \phi_q, z, z'),
\end{equation}
respectively.

Thanks to the cylindrical symmetry the Fourier transform of the Green's
function for an
arbitrary $\phi_q$ can be related to the one at $\phi_q=0$ through
\begin{equation}
   \tensor{G}(q_{\|}, \phi_q, z, z')
   =
   U(\phi_q)
   \tensor{G}(q_{\|}, \phi_q=0, z, z')
   [U(\phi_q)]^t,
\label{eq:fourierrotate}
\end{equation}
where
\begin{equation}
   U(\phi_q)
   =
   \left[
   \begin{array}{ccc}
      \cos{\phi_q} & -\sin{\phi_q} & 0 \\
      \sin{\phi_q} &  \cos{\phi_q} & 0 \\
      0 & 0 & 1
   \end{array}
   \right],
\end{equation}
and the superscript $t$ denotes transposition.
For $\phi_q=0$ 
$\tensor{G}^{p}(q_{\|},0,z,z')$
has four non-zero
components $xx$, $xz$, $zx$, and $zz$, while for
$\tensor{G}^{s}(q_{\|},0,z,z')$ only the $yy$ component is non-zero.

Likewise, for $\phi=0$ the Green's function in real space has 5 non-zero
components $xx$,  $yy$, $xz$, $zx$, and $zz$ and for a general $\phi$ we
have
\begin{equation}
   \tensor{G}(\rho, \phi, z, z')
   =
   U(\phi)
   \tensor{G}(\rho, \phi=0, z, z')
   [U(\phi)]^t.
\label{eq:realrotate}
\end{equation}

Therefore to calculate 
$\tensor{G}(\rho, \phi, z, z')$ in practice, we first calculate
$\tensor{G}$ for $\phi=0$ using the generalization of Eq.\
(\ref{eq:Fourier}) to the case of a layered material and then use Eq.\
(\ref{eq:realrotate}) to get the final result. The angular part of the
Fourier integral can be carried out analytically by making use of the
integral representation of
the Bessel functions 
\begin{equation}
   i^n J_n(z) = \frac{1}{2\pi} 
   \int_{0}^{2\pi} e^{i z \cos{\phi}} \cos{n\phi} \, d\phi.
\label{eq:bessel}
\end{equation}
We get
% \begin{widetext}
\begin{eqnarray}
   &&
   \tensor{G}(\rho, 0, z, z') 
   =
   \int
   \frac{d^2q_{\|}}{(2\pi)^2} 
   U(\phi_q)
   \tensor{G}(q_{\|}, 0, z, z')
   [U(\phi_q)]^t
   e^{i\vec{q}_{\|}\cdot\vec{r}_{\|}}
   \breakeq
   =
   \left[
   \begin{array}{ccc}
      G_{xx}(\rho,0,z,z') & 0 & G_{xz}(\rho, 0, z, z') \\
      0 & G_{yy}(\rho, 0, z, z') & 0 \\
      G_{zx}(\rho,0,z,z') & 0 & G_{zz}(\rho, 0, z, z') 
   \end{array}
   \right],
   \breakeq
\end{eqnarray}
and using 
$e^{i\vec{q}_{\|}\cdot\vec{r}_{\|}} = 
e^{iq_{\|}\rho \cos{\phi_q}}$ and Eq.\ (\ref{eq:bessel})
the different components are explicitly given by
\begin{eqnarray}
   &&
   G_{xx}(\rho, 0, z, z') 
   =
   \int_{0}^{\infty}
   \left[
      \left(
	 J_0(q_{\|}\rho)
	 -
	 \frac{J_1(q_{\|}\rho)}{q_{\|}\rho}
      \right)
   \right.
   \breakeq
   \left.
   \times
      G_{xx}(q_{\|},0,z,z')
	 +
	 \frac{J_1(q_{\|}\rho)}{q_{\|}\rho}
      G_{yy}(q_{\|},0,z,z')
   \right]
   \,
   \frac{dq_{\|}}{2\pi},
   \breakeq
\label{eq:Gxx}
\end{eqnarray}
\begin{equation}
   G_{xz}(\rho, 0, z, z') 
   =
   \int_{0}^{\infty}
	 iJ_1(q_{\|}\rho)
      G_{xz}(q_{\|},0,z,z')
   \,
   \frac{dq_{\|}}{2\pi},
\label{eq:Gxz}
\end{equation}
and
\begin{equation}
   G_{zz}(\rho, 0, z, z')
   =
   \int_{0}^{\infty}
	 J_0(q_{\|}\rho)
      G_{zz}(q_{\|},0,z,z')
   \frac{dq_{\|}}{2\pi}.
\label{eq:Gzz}
\end{equation}
The expression for $G_{yy}(\rho,0,z,z')$ is obtained by the index
replacements $xx\to yy$ and $yy\to xx$ in Eq.\ (\ref{eq:Gxx}), and
$G_{zx}(\rho,0,z,z')$ is obtained by replacing the index $xz$ in Eq.\
(\ref{eq:Gxz}) by $zx$.

The integrations in Eqs.\ (\ref{eq:Gxx}), (\ref{eq:Gxz}) and
(\ref{eq:Gzz}) nominally run along the real $q_{\|}$ axis, however, as
discussed in Ref.\ \onlinecite{Paulus:00} the numerical evaluation can be speeded
up by deforming the integration contour into the complex plane. 
As a first step we also divide the Green's function into two parts, the
homogeneous part $\tensor{G}_h$ that we have already discussed and given
explicit expressions for, and an indirect part
$\tensor{G}_{ind}$.\cite{Simsek:2010} 
The homogeneous part refers to the situation without interfaces; the
indirect part contains all contributions to the Greens function that at
any point involves the reflection or transmission of a wave at any of
the interfaces of the layered system. 
Both in real space and Fourier space 
\begin{equation}
   \tensor{G}
   =
   \tensor{G}_h
   +
   \tensor{G}_{ind}.
\label{eq:hominhom}
\end{equation}
Thus, in practice we evaluate $\tensor{G}_h$ from the explicit
expression in Eq.\ (\ref{eq:Ghexplicit}),\footnote{We define the homogeneous
Green's function to vanish identically, $\tensor{G}_h\equiv0$, whenever
the source and field points are in different layers.}
while the indirect part is
calculated from the Fourier integrals above. 
The functions in the respective integrals are
analytic in the lower half plane (LHP) but has two branch cuts in the upper
half plane (UHP) along the hyperbolas for which 
$\mathrm{Im}[p_1]=0$ 
and 
$\mathrm{Im}[p_N]=0$, respectively, where $p_1$ and $p_N$ are  given by 
Eq.\ (\ref{eq:pdef}) using the material properties of the top and bottom
layers. In addition the integrands may have one or several poles in the
UHP. We deform the integration contour so that it starts
from $q_{\|}=0$, first runs along the negative imaginary axis, then goes
parallel to the real axis until it reaches a point beyond the
singularities in the UHP where it goes back to the real axis. From there
the integration either proceeds along the real axis to values of
$q_{\|}$ large enough that further contributions are negligible, or
in the case that the lateral distance $\rho$ between the
source and field points is large a faster convergence is achieved by
rewriting the Bessel functions in terms of Hankel functions 
as \cite{Paulus:00}
\begin{equation}
   J_n(q_{\\}\rho) = \frac{1}{2} 
   \left[
      H_n^{(1)}(q_{\\}\rho) 
      +
      H_n^{(2)}(q_{\\}\rho) 
   \right],
\label{eq:Hankel}
\end{equation}
and then carrying out the integration of the $H_n^{(1)}$ part along a
vertical path in the UHP and the $H_n^{(2)}$ part along a vertical path
in the LHP.

For the purpose of (semi-)analytical calculations of the Green functions
it is often an advantage to deform the branch cuts, more about that
later.

\subsection{Reciprocity symmetry of the Green's function}

Reciprocity, which can be stated as ``interchanging the source and the field
probe does not change the result,'' is a central property of linear,
time-reversal-invariant electrodynamics. In our case this requires that
the Green's function fulfills the relation
\begin{equation}
   {G}_{\beta\alpha}(\vec{r}_2, \vec{r}_1) 
   =
   {G}_{\alpha\beta}(\vec{r}_1, \vec{r}_2).
\label{eq:reciproc}
\end{equation}
To show that this is in fact true one can go back to Eqs.\
(\ref{eq:Gpqformal}) and (\ref{eq:Gsqformal}), and first look at the
matrix appearing in the denominators,
$D^{\sigma}(z')$. 
While 
$D^{\sigma}(z')$
nominally appears to be a
function of the source coordinate $z'$,
it is in fact an invariant, independent of $z'$.
To see this we first note that $D^{\sigma}$ is independent of the
position of $z'$ within layer $l'$. Writing 
$D^{\sigma}$ as 
\begin{widetext}
\begin{equation}
   D^{\sigma}(z') = 
   \left[
      W^{+,\sigma}(d_{l'-1}-0,z_+)
   \right]^t
   S_{l'}^{+}(z', d_{l'-1})
   (-2ip_{l'} \sigma_z)
   S_{l'}^{-}(z', d_{l'})
   W^{-,\sigma}(d_{l'}+0,z_-),
\end{equation}
explicitly exposes the propagation in layer $l'$
and it is easy to show that
\begin{equation}
   S_{l'}^{+}(z', d_{l'-1})
   (-2ip_{l'} \sigma_z)
   S_{l'}^{-}(z', d_{l'})
   =
   S_{l'}^{+}(d_{l'}, d_{l'-1})
   (-2ip_{l'} \sigma_z)
   =
   (-2ip_{l'} \sigma_z)
   S_{l'}^{-}(d_{l'-1}, d_{l'}).
\label{eq:anywhere}
\end{equation}
It remains to see what happens to $D^{\sigma}(z')$ when $z'$ is moved
across an interface. We then have 
\begin{equation}
   D^{\sigma}(d_{l'}+0) = 
   \left[
      W^{+,\sigma}(d_{l'}+0,z_+)
   \right]^t
   (-2ip_{l'} \sigma_z)
   T_{l',l'+1}^{\sigma}
   W^{-,\sigma}(d_{l'}-0,z_-),
\end{equation}
just above an interface, and
\begin{equation}
   D^{\sigma}(d_{l'}-0) = 
   \left[
      W^{+,\sigma}(d_{l'}+0,z_+)
   \right]^t
   T_{l'+1,l'}^{\sigma}
   (-2ip_{l'+1} \sigma_z)
   W^{-,\sigma}(d_{l'}-0,z_-),
\end{equation}
just below.
But for both p and s polarization an explicit calculation shows that
\end{widetext}
\begin{equation}
   (-2ip_{l'} \sigma_z)
   T_{l',l'+1}^{\sigma}
   =
   T_{l'+1,l'}^{\sigma}
   (-2ip_{l'+1} \sigma_z).
\label{eq:anyside}
\end{equation}
Thus, Eqs.\ (\ref{eq:anywhere}) and (\ref{eq:anyside}) show that the
matrices $D^{p}$ and $D^{s}$ are invariant to all changes of $z'$, both
within a layer and from one layer to another.

To prove Eq.\ (\ref{eq:reciproc}) we 
reverse the  propagation direction in Eqs.\ (\ref{eq:Gpqformal}) and
(\ref{eq:Gsqformal}), which means that 
$\vec{q}_{\|} \to -\vec{q}_{\|}$,
$\tau \to -\tau$, 
$z$ and $z'$ and the layer indices $l$ and $l'$ are interchanged, and
$\hat{s}(\vec{q}_{\|}) \to \hat{s}(-\vec{q}_{\|}) = -\hat{s}(\vec{q}_{\|})$,
and
$\hat{p}^{\pm}(\vec{q}_{\|}) \to
\hat{p}^{\mp}(\vec{-q}_{\|})=\hat{p}^{\pm}(\vec{q}_{\|})$, 
and find that
\begin{equation}
   {G}_{\beta\alpha}^{\sigma}(-\vec{q}_{\|},z_2,z_1)
   =
   {G}_{\alpha\beta}^{\sigma}(\vec{q}_{\|},z_1,z_2)
\label{eq:reciprocFourier}
\end{equation}
for both p and s polarization.
As a consequence, inserting Eq.\ (\ref{eq:Gqsum}) into Eq.\
(\ref{eq:Fourier}) and substituting the integration variable 
$\vec{q}_{\|} \to -\vec{q}_{\|}$ 
we recover Eq.\ (\ref{eq:reciproc}) 
\begin{eqnarray}
   &&
   G_{\beta\alpha}(\vec{r}_2, \vec{r}_1) 
   =
   \int
   \frac{d^2q_{\|}}{(2\pi)^2}
   G_{\beta\alpha}(\vec{q}_{\|}, z_2, z_1)
   e^{i\vec{q}_{\|}\cdot(\vec{r}_{2\|}-\vec{r}_{1\|})}
   =
   \breakeq
   =
   \int
   \frac{d^2q_{\|}}{(2\pi)^2}
   G_{\beta\alpha}(-\vec{q}_{\|}, z_2, z_1)
   e^{i\vec{q}_{\|}\cdot(\vec{r}_{1\|}-\vec{r}_{2\|})}
   =
   G_{\alpha\beta}(\vec{r}_1, \vec{r}_2), 
   \breakeq
\end{eqnarray}
where we used Eq.\ (\ref{eq:reciprocFourier}) in the last step.

\subsection{Surface response from the Green's function}

As we have seen in this section, calculating $\tensor{G}$ for a
layered system is in general fairly involved, however, once it has been
calculated a lot of information can also be extracted from the Green's
function and its Fourier transform. As a first example we determine
the reflection factors for a plane wave impinging on one of the outer
interfaces at $z=d_1$ or $z=d_{N-1}$. 

For definiteness we concentrate on the reflection off the top interface
at $z=d_1$, and use  
\begin{equation}
   G_{\alpha\beta}(\vec{q}_{\|}, z, z') = \hat{\alpha} \cdot \vec{\cE},
\end{equation}
(cf.\ Eq.\ (\ref{eq:gkhom})) with a vector field $\vec{\cE}$ of the form shown
in Eq.\ (\ref{eq:vectorfield}) to find a relation
between the reflection coefficients and the Fourier transform of the
Green's function. We assume that the source is placed above the interface so
that the vector field  can be written in terms of the actual amplitudes we
introduced earlier
\begin{eqnarray}
   &&
   \vec{\cE} 
   =
   \left[
      \tensor{G}^{p}(\vec{q}_{\|}, d_1+0, z')
      +
      \tensor{G}^{s}(\vec{q}_{\|}, d_1+0, z') 
   \right]
   \cdot \hat{\beta}
   =
   \breakeq
   \hat{p}^{-} A_{1}^{-,p}(d_1)
   +
   \hat{p}^{+} B_{1}^{-,p}(d_1)
   +
   \hat{s}^{-} A_{1}^{-,s}(d_1)
   +
   \hat{s}^{+} B_{1}^{-,s}(d_1).
   \breakeq
\end{eqnarray}
Moreover, if the Green's function is divided into a homogeneous part and
an indirect part as in Eq.\ (\ref{eq:hominhom}), in
this case the terms with $A$ coefficients contribute to $\tensor{G}_{h}$
whereas the $B$ terms contribute to $\tensor{G}_{ind}$. We can therefore
conclude that the reflection coefficient for polarization $\sigma$ can
be written
\begin{equation}
   R_{\sigma} 
   =
   \frac{
      B_{1}^{-,\sigma}(d_1)
   }{
      A_{1}^{-,\sigma}(d_1)
   }
   =
   \frac{
      \hat{\sigma}^{+} \cdot 
      \tensor{G}_{ind}(\vec{q}_{\|}, d_1+0, z') \cdot \hat{\sigma}^{-}
   }{
      \hat{\sigma}^{-} \cdot 
      \tensor{G}_{h}(\vec{q}_{\|}, d_1+0, z') \cdot \hat{\sigma}^{-}
   }.
\end{equation}

\subsection{The Green's function and far-field calculations}

In a lot of situations one wants to calculate the scattered electric
field very far from the layered system. Given a source distribution
$j_{\mathrm{tot}}(\vec{r}')$ the field, retaining only non-zero terms in
Eq.\ (\ref{eq:lipp}), is
\begin{equation}
   E(\vec{r}) 
   =
   \int
   \tensor{G}(\vec{r}, \vec{r}') 
      i \omega \mu_0
      \vec{j}_{\mathrm{tot}}(\vec{r}') d^3r'.
\label{eq:Efar}
\end{equation}
The Green's function here can be written
\begin{eqnarray}
   \tensor{G}(\vec{r}, \vec{r}') 
   &&
   =
   \int
   \frac{d^2q_{\|}}{(2\pi)^2}
   \tensor{G}(\vec{q}_{\|}, z_+, z')
   \breakeq
   \times
   e^{i\vec{q}_{\|}\cdot(\vec{r}_{\|}-\vec{r}_{\|}')} 
   e^{i\sqrt{k_{B1}^2 - q_{\|}^2} (z-z_+)}
\end{eqnarray}
where $z_+$ lies above all layer interfaces $d_l$, as well as the source
$z'$. To get somewhat simpler expressions we first assume that we 
can set $z_+=0$. The integral can be
evaluated by the method of stationary phase which yields
\begin{equation}
   \tensor{G}(\vec{r}, \vec{r}') 
   =
   \frac{e^{ik_{B1}r}}{4\pi r} 
   e^{-ik_{B1}\sin{\theta}(x'\cos{\varphi}+y'\sin{\varphi})}
   \tensor{g}_{\mathrm{far}}(\theta,\varphi,z')
\label{eq:gasymp}
\end{equation}
where
\begin{equation}
   \tensor{g}_{\mathrm{far}}(\theta,\varphi,z')
   =
   -2ik_{B1}\cos{\theta}
   \tensor{G}(k_{B1}\sin{\theta}, \varphi, 0, z'),
\label{eq:gfar1}
\end{equation}
$r=|\vec{r}|$ and $k_{B1}=\sqrt{\varepsilon_1}\omega/c$.
In case $z_{+}>0$ one must use the generalized expression
\begin{equation}
   \tensor{g}_{\mathrm{far}}(\theta,\varphi,z')
   =
   -2ik_{B1}\cos{\theta} 
   e^{-ip_1z_+}
   \tensor{G}(k_{B1}\sin{\theta}, \varphi, z_{+}, z'),
\label{eq:gfar2}
\end{equation}
where the exponential function compensates for the propagation of the
outgoing wave included in $\tensor{G}$.

This expression for the far field is also useful in order to evaluate
the field generated in the layered system by an incident plane wave.
Assume that a plane transverse wave 
$$
   \vec{E}_{\mathrm{inc}} e^{i\vec{k}\cdot\vec{r}}
$$
impinges on the top surface of the stack of layers. 
This plane wave can be generated by a point source very far away at the
point $(r,\theta,\varphi)$ in spherical coordinates in the direction
where the wave comes from, i.e. 
\begin{equation}
   \vec{k} = (
      -k\sin{\theta}\cos{\varphi},
      -k\sin{\theta}\sin{\varphi},
      -k\cos{\theta}
   ).
\end{equation}
Comparison with Eq.\ (\ref{eq:Ghexplicit}) shows that
this requires a point source 
\begin{equation}
   \vec{j}(\vec{r}') 
   =
   \delta(\vec{r}'-\vec{r})
   \vec{E}_{\mathrm{inc}}
   \,
   \frac{4\pi r}{i\omega\mu_0}
   \, 
   e^{-ik_Br}
\label{eq:pointsource}
\end{equation}
at the point $\vec{r}$.
The full field at the point $\vec{r}_0=(x_0,y_0,z_0)$ 
in the layered structure can now be calculated
by inserting the source of Eq.\ (\ref{eq:pointsource}) in Eq.\
(\ref{eq:lipp}) (generalized to a layered background) and then applying
the reciprocity relation Eq.\ (\ref{eq:reciproc}), and Eq.\
(\ref{eq:gasymp}). This yields
\begin{equation}
   \vec{E}_0(\vec{r_0})
   =
   e^{-ik_{B1}\sin{\theta}(x_0\cos{\varphi}+y_0\sin{\varphi})}
   \left[
      \tensor{g}_{\mathrm{far}}(\theta,\varphi,z_0)
   \right]^t
   \vec{E}_{\mathrm{inc}}.
\label{eq:Ezero}
\end{equation}

\section{Long-range properties of the Green's function}
\label{sec:analytic}

The asymptotic behavior of the Green's function in the case that both
the source and field points lie close to a metal surface is a problem
that has attracted a lot of interest in the last few 
years.\cite{Aigouy:07,Leveque:07,Yang:Lalanne:09,Nikitin:PRL:10}

\begin{figure}[tb]
   \includegraphics[angle=0,width=8.0 cm]{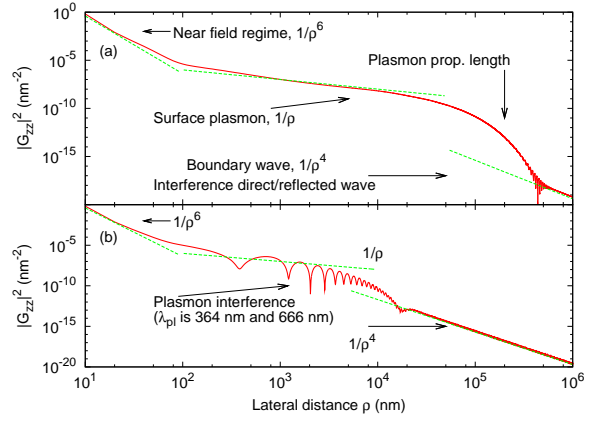}
   \caption{
      (color online)
      Behavior of the absolute square of the $zz$ element of the Green's
      function along the surface of (a) a gold sample, (b) a 20 nm thick
      gold film. The photon energy is 1.8 eV.
   }
\label{fig:asymptotic}
\end{figure}

Our results for the amplitude squared of the $zz$ element of the Green's
function are shown in Fig.\ \ref{fig:asymptotic} (which uses a
logarithmic scale on both axis). This figure
illustrates that the field generated by a point source near a metal film
and propagating outwards along the surface basically displays three
different regimes: At short range from the source the dipole field
originating from the source (and its image) dominates completely
giving rise to a large field that however drops off as $1/\rho^6$. After
that follows a range of distances where plasmon propagation along the
surface gives the dominant contribution to the Green's function. 
The plasmons are cylindrical waves confined to the metal surface so
their amplitudes decay as $1/\sqrt{\rho}$ in the absence of power losses. 
A Au film on a glass substrate supports  two different plasmons, and in
this case we see interference between them.
Eventually, once we get to lateral distances between the source and
field points corresponding to the plasmon propagation length
the Green's function drops exponentially due to losses in
the metal film and we enter the domain where the main contribution comes
from a boundary wave, also known as a Norton wave, \cite{Nikitin:PRL:10}
propagating along the interface. 

% Mention interference

Naively one may expect $|G_{zz}|^2$ to decay as $1/\rho^2$ in the boundary
wave regime, but in fact one finds a faster decay, $\sim 1/\rho^4$. 
Refs.\  \onlinecite{Aigouy:07,Leveque:07,Yang:Lalanne:09,Nikitin:PRL:10}
present a number of derivations of this behavior. The basic physical
reason behind the rapid decay is a destructive interference between the
direct wave emerging from the source, and the wave reflected off the
surface.  As is easily seen from the expressions in Eq.\
(\ref{eq:Fresnelf}), exactly at grazing incidence (where $p_l=0$) 
both of the Fresnel
reflection coefficients $f^s$ and $f^p$ equal -1, which means that to
lowest order the sum of incident and reflected wave vanishes. 
Away from grazing incidence the incident and reflected wave do not
cancel exactly and what remains (with $|G_{zz}|^2\sim1/\rho^4$)
is the result of the interference between these contributions.

\begin{figure}[tb]
   \includegraphics[angle=0,width=8.0 cm]{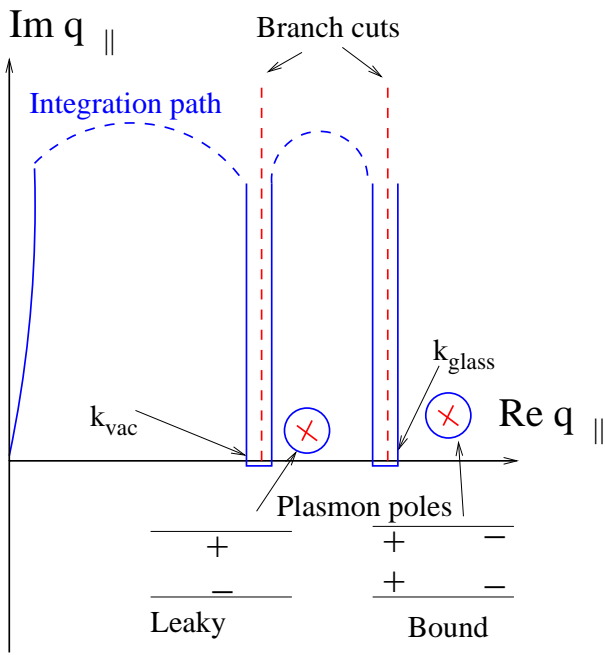}
   \caption{
      (color online)
      Deformation of the integration path in the complex plane in order
      to capture the asymptotic behavior of the Green's function along
      the surface of the metal film. 
   }
\label{fig:contour}
\end{figure}

% Go on discussing what it looks like in the complex plane.
In order to understand and calculate the asymptotic behavior of the
Green's function it is useful to study the behavior of the Fourier
transform $\tensor{G}({q}_{\|},0, z,z')$ in the complex plane. Figure
\ref{fig:contour} shows the general structure for the case of a
three-layer structure vacuum/metal/dielectric. The FT has two branch
points at $k_{B1}$ and $k_{B3}$, the vacuum and dielectric wave numbers.
The physical branch cuts, as discussed in connection with Eq.\
(\ref{eq:pdef}), in this case follows the real axis back to the origin
and then runs out along the positive imaginary axis. For the purpose of
performing the Fourier integral for large separations between the source
and field points it is, however, better to deform the branch cuts to run
parallel with the imaginary as shown in Fig.\ \ref{fig:contour}.

Now the Bessel function in the integral Eq.\ (\ref{eq:Gzz}) can be split
into two Hankel functions as indicated in Eq.\ (\ref{eq:Hankel}). The integral
containing $H_n^{(2)}(q\rho)$ is calculated by deforming the contour to run
far into the LHP, which for large $\rho$ yields negligible
contributions. The integral containing $H_{n}^{(1)}(q\rho)$ is
calculated by deforming the contour to run far into the UHP where this
Hankel function is exponentially small so that the contributions from
most of the contour are negligible. However, unlike in the LHP, the
contour in the upper half plane must return to the real axis (or its
vicinity) at every obstacle in the form of a branch cut or pole, and
these parts of the integration path yield the dominating contributions
to the Green's function for large $\rho$. \cite{Miller:2006}
The singularities closest to the real axis gives the contributions with
the farthest range in $\rho$.

For the case illustrated in Figs.\ \ref{fig:asymptotic} and
\ref{fig:contour} the contributions from the integration along the
branch cuts give the long-range, boundary wave contribution that
persists over the entire range of distances in Fig.\
\ref{fig:asymptotic}.  A close look at the result for a film 
shows that the long-range tail exhibits some small oscillations between
the contributions from the two different boundary waves, the one at the
vacuum-gold interface and the one at the gold-glass interface.

As already said the boundary waves become the dominant contribution for
values of $\rho$ beyond the plasmon propagation length. In Fig.\
\ref{fig:asymptotic}(b) this happens around $\rho\sim10 \unit{\mu m}$. The
behavior for $\rho \sim$ 1--10 $\unit{\mu m}$ is dictated by two different
plasmons corresponding to the two poles in the UHP. The pole to the
right of both branch cuts corresponds to a charge-symmetric, bound 
plasmon with a wavelength of
364 nm, shorter than the wavelengths of $\hbar \omega=1.8 \unit{eV}$
light in both glass and vacuum. 
\footnote{In this case, with different dielectric environments on each
side of the metal film, the mode is of course not completely symmetric with
respect to the surface charges.}
This mode is thus bound to the film, 
evanescent in both vacuum and glass, and the pole lies on the physical
sheet of the Riemann surface. The other plasmon with a wave\-length 666
nm is evanescent in vacuum, but propagating in glass. Therefore this
charge-asymmetric mode is termed leaky since energy is lost to the glass
side.\cite{Stegeman:1986} 
From a formal point of view this mode does not correspond to a
bound state and the corresponding pole lies on a higher sheet of the
Riemann surface, but is brought out in the open by the contour
deformation.

\section{Solution of the scattering problem}
\label{sec:solution}

The ultimate goal with the use of the Green's function method is of course
to solve electrodynamics problems in the form of Eq.\  (\ref{eq:lipp})
generalized to the situation with a layered background,
\begin{eqnarray}
   &&
   \vec{E} (\vec{r})
   =
   \frac{\tensor{L}\cdot\vec{j}(\vec{r})}{i\omega\varepsilon_0\varepsilon_B(\vec{r})}
   -
   \frac{\Delta \varepsilon(\vec{r)}}{\varepsilon_B(\vec{r})} 
   \tensor{L} \cdot \vec{E}(\vec{r})
   \breakeq
   +
   \int_{V_j-V_{\delta}}
   \tensor{G}(\vec{r}, \vec{r}')
   \left(
      i \omega \mu_0
      \vec{j}(\vec{r}')
      +
      k_0^2 \Delta \varepsilon(\vec{r'}) \vec{E}(\vec{r'})
   \right)
   d^3r'.
   \breakeq
\label{eq:lipp2}
\end{eqnarray}
We focus on the case where the current sources $\vec{j}$ are well
separated from the scatterers so that the contribution from the current
term in the integral in Eq.\ (\ref{eq:lipp}) can be written as an
incident field,
\begin{equation}
   \vec{E}_0(\vec{r}) 
   =
   \int_{V_j-V_{\delta}}
   \tensor{G}(\vec{r}, \vec{r}')
      i \omega \mu_0
      \vec{j}(\vec{r}')
   d^3r'.
\label{eq:E0def}
\end{equation}
If the sources are very far away we have a situation where a plane wave
is incident on the layered system, and gives rise to reflected and
transmitted waves in the system, all of this is dictated by the presence
of the Green's function in Eq.\ (\ref{eq:E0def}). With an incident field
written as $\vec{E}_{\mathrm{inc}}e^{i\vec{k}\cdot\vec{r}}$ we obtain a
driving field $\vec{E}_0(\vec{r})$ in the layered system in accordance with Eq.\
(\ref{eq:Ezero}).

Then Eq.\ (\ref{eq:lipp2}) can be written
\begin{widetext}
\begin{equation}
   \vec{E} (\vec{r})
   =
   \vec{E}_0 (\vec{r})
   -
   \frac{\Delta \varepsilon(\vec{r)}}{\varepsilon_B(\vec{r})} 
   \tensor{L} \cdot \vec{E}(\vec{r})
   +
   \int_{V_j-V_{\delta}}
   \tensor{G}(\vec{r}, \vec{r}')
      k_0^2 \Delta \varepsilon(\vec{r'}) \vec{E}(\vec{r'})
   d^3r'.
\label{eq:lipp3}
\end{equation}

By moving all terms involving the full field $\vec{E}(\vec{r})$ to the
left-hand side, thus leaving only the driving field $\vec{E}_0(\vec{r})$
on the right hand side, and then discretizing the electric field on a
mesh of equally sized cubic elements that covers all scatterers we arrive at 
\begin{equation}
   \vec{E}_{qrs}
   +
   \frac{\Delta \varepsilon_{qrs}}{\varepsilon_{B,qrs}} 
   \tensor{L} \cdot \vec{E}_{qrs}
   -
   k_0^2 \Delta \varepsilon_{qrs} 
   \tensor{M}
   \vec{E}_{qrs}
   -
   {\sum_{q'r's'}}^{'}
   k_0^2 \Delta \varepsilon_{q'r's'} V_M 
   \tensor{G}_{q-q',r-r',s,s'} 
   \vec{E}_{q'r's'}
   =
   \vec{E}_{0,qrs}.
\label{eq:lipp4}
\end{equation}
\end{widetext}
Here $q$, $r$, and $s$ are discrete coordinates for the mesh elements in
the $x$, $y$, and $z$ directions, respectively. 
With a mesh side $a_M$ and an equivalent radius $R_M$ the volume of a
mesh element is
\begin{equation}
   V_M = a_M^3 =
   \frac{4\pi R_M^3}{3}.
\end{equation}
The term containing $\tensor{M}$ describes self interaction within a
mesh element. We use an approximation for $\tensor{M}$ corresponding to
a spherical mesh element of radius $R_M$,
\begin{equation}
   \tensor{M}
   =
   \frac{2}{3k_0^2}
   \left[
      (1-ik_0R_M) \exp(ik_0R_M) - 1
   \right]
   \tensor{1}.
\end{equation}
At the same time the self-interaction term is excluded from the sum (as
indicated by the prime) since including this would involve singular
contributions to the Green's function.

Equation (\ref{eq:lipp4}) corresponds to a system of 
linear equations; the left hand side can be seen as a $3N_M\times3N_M$ matrix
multiplying a vector with $3N_M$ elements, $N_M$ being the total number of
mesh elements. 
We solve the system of equations iteratively using the stabilized
biconjugate gradient method, BiCGstab(2).\cite{Sleijpen:93,*Sleijpen:94}
The iterative solution involves a large number of matrix
multiplications. The contribution from the term in which the Green's
function multiplies the electric field is, as can be seen in Eq.\
(\ref{eq:lipp4}), the result of a convolution sum in the $x$ and $y$
directions. This means that the matrix multiplication can be speeded up
by using a Fast Fourier transform (FFT) in these two
directions.\cite{Golub:1996, *NumRec} The same technique is used in the
DDA method.\cite{Draine:94}
We calculate the Fourier transforms of the
Green's function and the electrical field and multiply the transforms by
each other locally on the mesh in Fourier space and then transform the
product back to the real space mesh.
The use of the FFT is crucial in reducing computation times, since
most of the computational effort required in determining the
electric field in the scattering volume goes into solving the equation
system, thus essentially the repeated matrix multiplications. 
The calculation of the Green's function, on the other hand, just needs
to be done once per photon frequency, combination of $z$ and $z'$, and
in-plane distance $\rho$.

Once we have a converged solution to the system of equations the
electric field inside the scatterers is known. At this point Eq.\
(\ref{eq:lipp3}) provides an explicit expression for the electric field
everywhere else in space that can be evaluated by discretizing the
integral as in Eq.\ (\ref{eq:lipp4}). 

The results presented in the next section focuses on the scattering
cross section and thus depend on the far field which can be found from a
discretized version of Eq.\ (\ref{eq:Efar}) using Eqs.\
(\ref{eq:gasymp}) and (\ref{eq:gfar2}).
The scattering cross section is given by
\begin{equation}
   \frac{d\sigma}{d\Omega}
   =
   \frac{r^2 S_r}{S_{in}},
\end{equation}
where $S_r$ is the radial component of the Poynting vector at a large
distance $r$ from the scatterers and $S_{in}$ is the Poynting vector of
the incident field. Given the (transverse) far field $\vec{E}(\vec{r})$,
\begin{equation}
   S_r
   =
   \frac{1}{2} \,c \, \varepsilon_0 \,
   \sqrt{\varepsilon_B} 
   |\vec{E}(\vec{r})|^2,
\end{equation}
where $\varepsilon_B$ is the dielectric function of the 
material the radiation is scattered into.

\section{Scattering off nanoholes in a thin metal film}
\label{sec:holes}

We now turn to calculating scattering spectra off nanoholes in a thin Au
film. Such systems have been studied experimentally by Rindzevicius
{\it et al.}\cite{Tomas:hole:07} and Alaverdyan {\it et
al.}\cite{Alaverdyan:hole:natphys:07}

\begin{figure}[htb]
   \includegraphics[angle=0,width=8.0 cm]{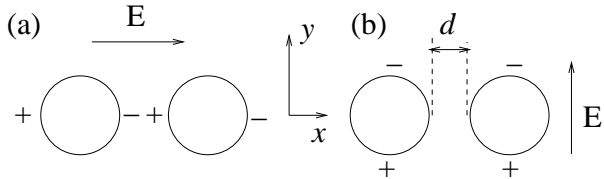}
   \caption{
      Illustration of illumination of two nanoholes with an electric
      field parallel to the dimer axis [in (a)], and perpendicular to
      the dimer axis [in (b)]. We also indicate how charges in the metal
      film surrounding the holes will be distributed in the case that
      the frequency of the incident light lies well below the
      single-hole resonance. One should note that the behavior of
      nanoholes in terms of induced charges is essentially opposite to
      that of metallic nanoparticles.
   }
\label{fig:polarize}
\end{figure}

To study the problem theoretically we let a number of circular
cylindrical holes in a Au film on top of a glass substrate act as
scatterers. The Au film here has the same thickness, 20 nm, as in the
experimental studies. % more geometry
The system is driven by a plane wave that impinges  on the film (and
the holes) at normal incidence, polarized either parallel to the symmetry
axis of the hole chain or perpendicular to that symmetry axis, as
illustrated in Fig.\
\ref{fig:polarize}.  We study primarily the forward scattering
cross section as the edge-to-edge distance $d$, between the holes is
varied.

\begin{figure}[htb]
   \includegraphics[angle=0,width=8.0 cm]{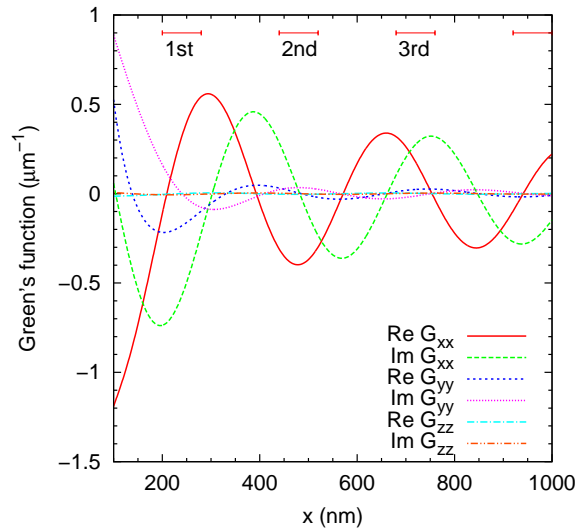}
   \caption{
      (color online)
      The diagonal components of the Green's function as a function of
      the lateral distance $x$ along a 20 nm thick Au film on a glass
      substrate. Both the source and field
      points are placed in the middle of the film to best describe
      hole-hole interaction. The marks near the upper border show where
      the 1st, 2nd, and 3rd neighbor hole is placed in a chain of holes
      with a radius of 40 nm and edge-to-edge distance $d=160
      \unit{nm}$. (We stress, though, that the Green's function here has been
      calculated in the absence of any holes.)
   }
\label{fig:gofx}
\end{figure}

As a prelude we look at the 
Green's function with both the source and field points placed inside
the metal film, which is a central quantity determining the interaction between
different nanoholes in the film. Figure \ref{fig:gofx} displays the
behavior of the diagonal elements of $\tensor{G}$ for the vacuum/Au
film/glass substrate system at a representative
photon energy of
1.8 eV ($\lambda\approx 690 \unit{nm}$) as a function of the
lateral separation $x$.
The $xx$ element is by far the strongest over most of the
range of distances $x$, between the source and field points. A source
pointing in the $x$ direction can excite plasmons propagating in the $x$
direction which explains why we have long-range interactions in this
case. These plasmons are of the bound, charge-symmetric type discussed
in Sec.\ \ref{sec:analytic}, and illustrated in Fig.\ \ref{fig:contour}.
The bound plasmon wavelength for $\hbar\omega=1.8 \unit{eV}$ is
$\approx 364 \unit{nm}$.
The $yy$ element is of comparable strength as $G_{xx}$ for
distances up to $\sim 200 \unit{nm}$, i.e.\ in the near-field zone. However, for
larger $x$ the $yy$ element is much smaller because a dipole pointing in
the $y$ direction cannot excite plasmons propagating in the $x$
direction. Finally looking at $G_{zz}$ we see that this component is
much smaller than $G_{xx}$ for all $x$ values. 
This is due to the boundary conditions for the electric field at a metal
interface, which strongly suppress the normal component inside the
metal. As a consequence of reciprocity this also means that a source
inside the metal film oriented perpendicular to the interfaces is not
very effective in generating electric fields elsewhere.

\begin{figure}[htb]
   \includegraphics[angle=0,width=8.0 cm]{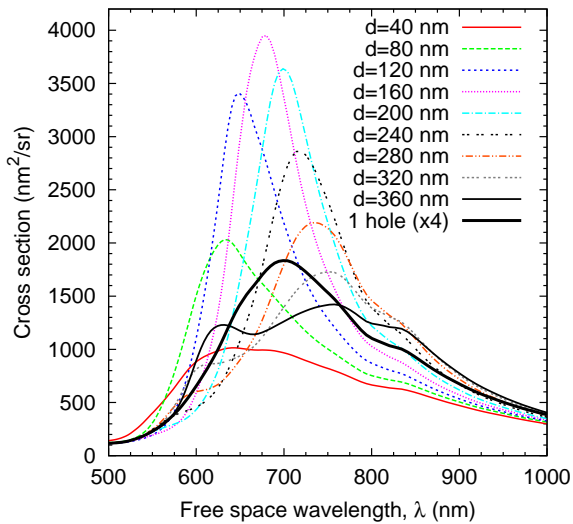}
   \caption{
      (color online)
      Calculated forward scattering spectra for two nanoholes of
      diameter 80 nm in a 20 nm thick Au film placed on a glass
      substrate illuminated at normal incidence by light polarized
      parallel to the dimer axis. The different curves show results for
      a series of edge-to-edge distances $d$ between the holes as
      indicated in the key. The curve marked ``1 hole'' shows the
      corresponding result, adjusted for the scattering volume, for the
      case of a single hole.
   }
\label{fig:parallell2}
\end{figure}

\begin{figure}[htb]
   \includegraphics[angle=0,width=8.0 cm]{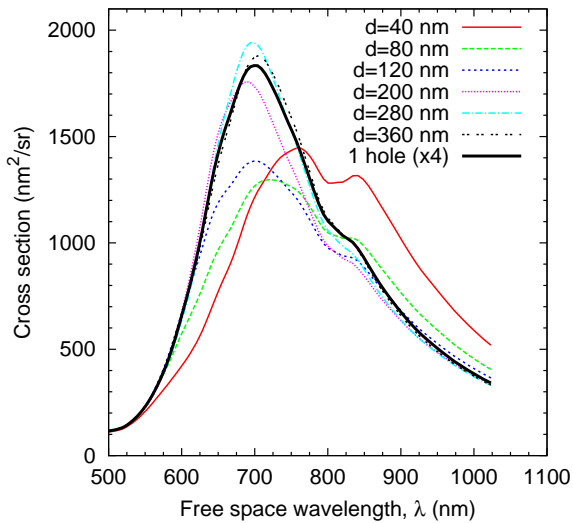}
   \caption{
      (color online)
      Forward scattering spectra for two nanoholes as in Fig.\
      \ref{fig:parallell2}, however, here the incident light is polarized
      perpendicular to the dimer axis.
   }
\label{fig:perpendicular2}
\end{figure}

The behavior of different elements of the Green's function
leads to differences in the hole-hole interaction depending on the
polarization direction of the incident light (illustrated in  
Fig.\ \ref{fig:polarize});
interaction effects are much more important in the case of parallel
polarization.
The consequences are clearly seen
in Figs.\ \ref{fig:parallell2} and \ref{fig:perpendicular2}. Figure
\ref{fig:parallell2}, to begin with, shows calculated scattering
cross sections for two nanoholes of diameter 80 nm that are illuminated
by light polarized parallel to the dimer axis. Each curve
corresponds to a different edge-to-edge separation between the holes. To
make a comparison that brings out the effect of hole-hole interactions
the result for a single hole is also shown. 
This result is multiplied by 4 to adjust to the
difference in scattering volume between the one- and two-hole cases.
For a small separation between the holes the scattering
cross section is suppressed and blue-shifted compared with the one-hole
case. This is a result of $G_{xx}$ being negative for $x$ smaller than
$\approx 200 \unit{nm}$ (an edge-to-edge separation of 40 nm corresponds
to a center-to-center distance of 120 nm). The shift can also be
understood in view of Fig.\ \ref{fig:polarize}(a): the figure shows that
for frequencies below resonance the field caused by the induced charges
at one hole
will counteract the external field at the other hole, but this situation
is reversed for frequencies above the single-hole resonance hence the
blue-shift.
With an increasing distance between the holes the scattering cross
section increases and its maximum red-shifts, and a maximum in the cross
section occurs for $d=160 \unit{nm}$, corresponding to a distance of 240
nm between the hole centers. As can be seen in Fig. \ref{fig:gofx} this
is close to the distance where $\mathrm{Re}[G_{xx}]$ has a maximum. In
this situation there is a constructive interference at one hole between
the incident field and the field scattered off the other hole.
For larger $d$ the scattering cross section continues to red-shift,
while the peak value falls off. For the largest separation $d=360
\unit{nm}$, we in fact see a new peak building up at the blue end of the
spectrum (near 650 nm). For even larger $d$ this peak grows and
red-shifts reaching a second maximum around $d=560 \unit{nm}$
corresponding to  a
center-to-center separation right near the
second maximum of $\mathrm{Re}[G_{xx}]$ in 
Fig. \ref{fig:gofx} at $x\approx650 \unit{nm}$.
In Ref.\ \onlinecite{Alaverdyan:hole:natphys:07} it is argued that the
scattering from a chain of holes should show maxima whenever an odd
number of half surface plasmon wavelengths can be fit in between two
holes. We note that in the present case with $\lambda_{\mathrm{pl}}=364
\unit{nm}$, this predicts scattering maxima for
$d=\lambda_{\mathrm{pl}}/2=182 \unit{nm}$ and
$d=3\lambda_{\mathrm{pl}}/2=546 \unit{nm}$, which indeed agrees very
well with the calculated results. Still, looking at the
behavior of the Green's function is a more general way of predicting
resonance conditions. 
% Maybe something more here

The results in Fig. \ref{fig:perpendicular2} calculated with the
incident light polarized perpendicular to the hole dimer axis show much
less variation with $d$. There is a suppression and a red-shift of the
cross section for $d=40 \unit{nm}$. This is expected given the
basic behavior illustrated in Fig.\ \ref{fig:polarize}(b) since in this
case the field from the induced charges acts to enhance the external
field at frequencies below the single-hole resonance.
But with increasing $d$ the two-hole result rather quickly approaches
the adjusted one-hole result, i.e.\ the spectrum is only marginally
affected by hole-hole interactions. This can be anticipated by a look at
the results for $G_{yy}$ in Fig.\ \ref{fig:gofx} which shows that the
long-range interaction is rather weak for this configuration.

\begin{figure}[htb]
   \includegraphics[angle=0,width=8.0 cm]{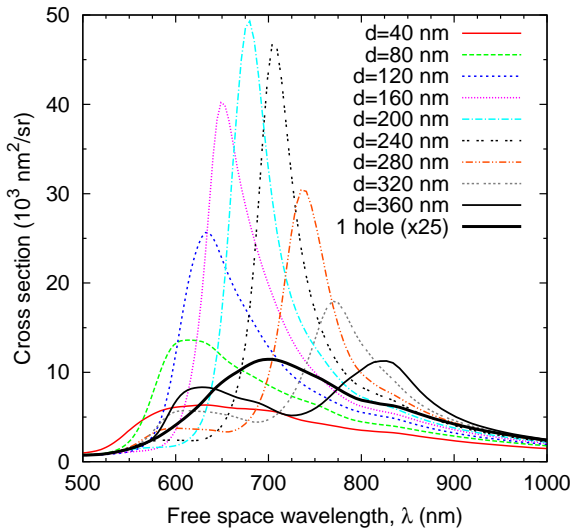}
   \caption{
      (color online)
      Forward scattering spectra for 5 nanoholes illuminated by light
      polarized along the axis of the chain of holes.
      The remaining parameters are the same as in Fig.\
      \ref{fig:parallell2}.
   }
\label{fig:parallell5}
\end{figure}

\begin{figure}[htb]
   \includegraphics[angle=0,width=8.0 cm]{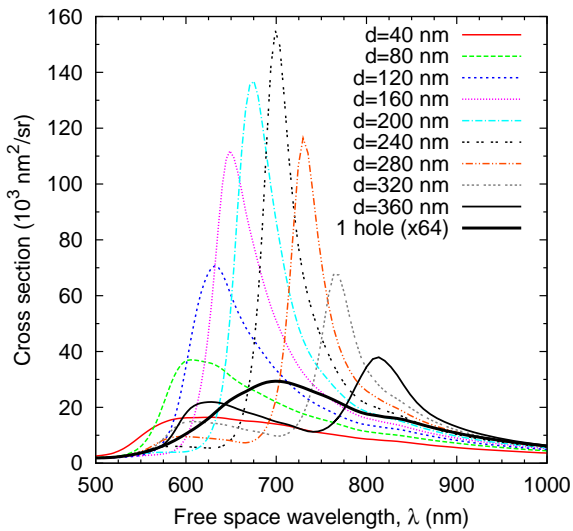}
   \caption{
      (color online)
      Forward scattering spectra for 8 nanoholes illuminated by light
      polarized along the axis of the chain of holes.
      The remaining parameters are the same as in Fig.\
      \ref{fig:parallell2}.
   }
\label{fig:parallell8}
\end{figure}

Figures \ref{fig:parallell5} and \ref{fig:parallell8} show scattering
spectra for chains of 5 and 8 holes, respectively, illuminated by light
polarized along the axis of the chain. These results show the same
trends as those in Fig.\ \ref{fig:parallell2}, but one can still make
some additional observations. (i) The fact that we have more holes means
that the collective effects of hole-hole interactions are stronger since
the holes inside the chain now have two nearest neighbors. Consequently
the peak position shifts more now when changing $d$ and the
spectra rise higher above the (adjusted) one-hole result.
(ii) The maximum
scattering cross section is obtained at somewhat larger values of $d$ 
compared with the two-hole case.
The reason is that
not only nearest-neighbor interactions matter now. The cross section can be
increased by moving the next-nearest neighbor hole closer to
the second maximum of 
$\mathrm{Re[G_{xx}]}$ at $x=650 \unit{nm}$, see Fig.\ \ref{fig:gofx},
something that is achieved by an increase of $d$.
(iii) 
We also see that the spectral
features are sharper here than in the two-hole case. 
This is a rather natural consequence of the
facts discussed above. An increasing number of holes brings an
increasing degree of collective behavior and constructive interference
to the optical response of the hole system, which at the same time is
more sensitive to changes in either the photon energy of the incident
light, the hole-hole separation, or for that matter the dielectric
environment. 
Going from 2 holes to 5
makes more of a qualitative difference than increasing the number from
5 to 8. The reason for this is primarily the fact that nearest-neighbor
interactions play a dominant role; with two holes in the chain both of
them just have one nearest neighbor, whereas for 5- or 8-hole chains the
majority of the holes have two nearest neighbors.

The results presented here agree very well with the experimental results
found in Ref.\ \onlinecite{Alaverdyan:hole:natphys:07}, see in particular
Fig.\ 2 there. (i) As in the experiment nanohole interactions play an
important role when the electric field is polarized along the axis of
the hole chain, while interactions only have a minor influence on the
spectrum in the case of perpendicular polarization. (ii) For parallel
polarization the experimental scattering spectrum goes through the same
development as in Figs.\ \ref{fig:parallell2}, \ref{fig:parallell5}, and
\ref{fig:parallell8}. For the smallest edge-to-edge distances the
spectrum is suppressed and blue-shifted, but as $d$ increases a strong
successively red-shifted peak builds up. (iii) The maximum scattering
cross section in the two-hole case is reached for $d=160 \unit{nm}$
here, and for $d=150 \unit{nm}$ in the experiment, see Fig. 2(a) of
Ref.\ \onlinecite{Alaverdyan:hole:natphys:07}. These peak wavelengths
differ somewhat, $\approx 655 \unit{nm}$ in the experiment and $\approx
675 \unit{nm}$ here, part of the reason for this is probably 
that the holes used in the experiment were somewhat smaller, with a
diameter of 75 nm.  (iv) Comparing the experimental results for 8 holes
with those for 2 holes [Fig 2(a) of the experimental paper] we also see
much the same trends as discussed above. More holes give stronger and
sharper peaks and bigger wavelength shifts as a function of the
edge-to-edge distance, just as in the calculation.

\section{Summary}
\label{sec:summary}

In this paper we have presented a derivation of the electromagnetic
Green's function in systems where the background dielectric function
varies stepwise along one of the coordinate directions, $z$. The
derivation is built on a transfer-matrix calculation of the Fourier
transform of the GF. We have discussed certain symmetry properties of
the Green's function and also studied its long-range properties in real
space based on the analytic properties of the Fourier transform in the
complex plane. 

As an example of an application we have studied the long-range
properties of the Green's function near a thin Au film on a glass
substrate. We find there three different regimes depending on the
lateral distance $\rho$ between the source and field points: (i) A near-field
regime where the the square of the GF decay as $1/\rho^6$. (ii) For 
$100 \unit{nm} \alt \rho \alt 10 \unit{\mu m}$ the Green's function is
dominated by contributions from propagating surface plasmons, and
$|G|^2\sim1/\rho$. (iii) Finally, for larger distances, beyond the
surface plasmon propagation length, the Green's function is dominated by
contributions from boundary waves (Norton waves) grazing the interface. 
A nearly destructive
interference between the incident and reflected wave results in the
intensity $\propto |G|^2$ decaying as $1/\rho^4$ in this case.

We have also applied the Green's function method to a calculation of the
scattering off two or several nanoholes in a thin Au film. We find a
strong hole-hole interaction mediated by the surface plasmons of the Au
film provided that the incident electric field is polarized along the
axis of the hole chain. By increasing the number of holes the scattering
spectrum gets sharper features and becomes more sensitive to changes in
geometry, photon energy, or dielectric environment, something that can
have applications in for example biochemical sensing.

\section*{Acknowledgments}

I have benefited from discussions with Andreas Thore, Vladimir
Miljkovic, Mikael K\"all, and Peter Apell.
Financial support from the Swedish Research Council (VR) is gratefully
acknowledged.

\end{document}